\documentclass[conference]{IEEEtran}

\usepackage{amsmath,amsfonts}
\usepackage{amssymb}
\usepackage{enumerate,letltxmacro}
\usepackage{graphicx}
\usepackage{cite}
\usepackage{subfigure}
\usepackage{multirow}
\usepackage{algorithm}
\usepackage{algorithmic}
\usepackage{mathtools}
\usepackage{hyperref}
\usepackage{verbatim}
\usepackage{xcolor}

\hypersetup{draft}

\IEEEoverridecommandlockouts

\newtheorem{theorem}{Theorem}

\newtheorem{corollary}{Corollary}
\newtheorem{proposition}{Proposition}

\title{On Characterization of Entropic Vectors at the Boundary of Almost Entropic Cones}
\author{\IEEEauthorblockN{Hitika Tiwari and Satyajit Thakor}
\IEEEauthorblockN{School of Computing and Electrical Engineering\\
Indian Institute of Technology Mandi, Himachal Pradesh, India}
email: T17144@students.iitmandi.ac.in, satyajit@iitmandi.ac.in
}

\begin{document}
\maketitle
\begin{abstract}
The entropy region is a fundamental object in information theory. An outer bound for the entropy region is defined by a minimal set of Shannon-type inequalities called elemental inequalities also referred to as the Shannon region. This paper focuses on characterization of the entropic points at the boundary of the Shannon region for three random variables. The proper faces of the Shannon region form its boundary. We give new outer bounds for the entropy region in certain faces and show by explicit construction of distributions that the existing inner bounds for the entropy region in certain faces are not tight. 
\end{abstract}
\section{Introduction}
A vector (or a point) $\mathbf{h}=[h_\beta: \emptyset\neq \beta \subseteq [n]]^\intercal,$ where $[n]=\{1,2, \ldots,n\}$, in $\mathbb{R}^{2^n-1}$ is said to be \textit{entropic} if there exists a collection of $n$ discrete random variables $\{Y_1,\ldots,Y_n\}$ over finite support sets such that $h_\beta$\footnote{For simplicity, the set $\beta$ in the subscript is represented without braces and by juxtaposition of its elements, e.g., for $\beta=\{i,j\}$, $h_{\beta}$ is denoted by $h_{ij}$.} is equal to the entropy function $H(Y_\beta)$ for all $ \emptyset\neq \beta \subseteq [n]$ where $Y_{\beta}\triangleq\{Y_i:i\in\beta\}$. In this context, the Euclidean space $\mathbb{R}^{2^n-1}$ is referred to as the \textit{entropy space}.
The \textit{entropy region} is denoted by $\Gamma_n^*$ and defined as \cite{Yeu97}
\begin{align}\label{eq:entropyR}
\Gamma_n^*\triangleq \{\mathbf{h} : \mathbf{h} \text{ is entropic}\}.
\end{align}

Any Shannon information measures can be represented as a sum of the following two elemental forms \cite{Yeu97}.
\begin{align}
H(Y_{\beta}\mid Y_{\beta^c}),& \quad|\beta|=1\\
I(Y_{i}; Y_{j}\mid  Y_{\beta}),& \quad i\neq j \text{ and }\beta\subseteq \{i, j\}^c
\end{align}
Non-negativity of elemental forms are called elemental inequalities. These inequalities imply all Shannon-type inequalities \cite{Yeu97}. The \textit{Shannon region} is 
defined as
\begin{align}\label{eq:ShannonR}
\Gamma_n\triangleq \{\mathbf{h} : \mathbf{h} \text{ satisfies elemental inequalities}\}.
\end{align}

Any entropic $\mathbf{h}$ must satisfy Shannon-type inequalities and hence
$\Gamma_n^*\subseteq \Gamma_n.$
The regions $\Gamma_n$ and  $\overline{\Gamma}^*_n$ (the closure of $\Gamma^*_n$, also referred to as the \textit{almost entropic region}) are cones \cite{Yeu97,Yeu08}. In particular, $\Gamma_n$ is a pointed polyhedral cone whereas $\overline{\Gamma}^*_n$ is a pointed convex cone (not polyhedral in general). The entropy region $\Gamma^*_n$ is a non-convex set in general. For the cases of $n=1,2$ the entropy region and Shannon bound are the same. For $n=3$, $\Gamma^*_3 \subsetneq \Gamma_3$ but the closure of $\Gamma^*_3$, denoted $\overline{\Gamma}^*_3$, is the same as $\Gamma_3$, i.e.,
\begin{align}
\overline{\Gamma}^*_3=\Gamma_3.
\end{align}
Thus, the Shannon region for three random variables is an almost entropic cone and there exist non-entropic points at the boundary of the Shannon region \cite{ZhaYeu97}. The boundary can be viewed as the set of points in the proper faces of the cone. In this paper, we focus on characterization of the entropy regions in the faces of the Shannon bound for three random variables.
	
In Section \ref{sec:Prelim}, faces of $\Gamma_3$ containing non-entropic points and known results on characterization are discussed. In Section \ref{sec:Main Results}, we present outer bounds (converse-type results) on the entropy region for seven faces of $\Gamma_3$. We also show that existing inner bounds for two faces are loose via explicit construction of distributions with corresponding entropic points in the faces. The approach used can be applied to investigate tightness of known inner bounds for constrained entropy regions or to obtain distributions corresponding to entropic points in a constrained entropy region. Finally, conclusion is presented in Section \ref{sec:Conclusion}.

\section{Preliminaries}\label{sec:Prelim}
For 3 random variables, let a vector $\mathbf{h}$ in the entropy space be denoted
$$\mathbf{h}=[h_1,h_2,h_3,h_{12},h_{13},h_{23},h_{123}]^\intercal.$$
Replacing an inequality of an elemental form with equality and then taking its intersection with $\Gamma_3$, we obtain a $2^n-2=6$ dimensional face called a \textit{facet}. A lower dimensional face is a subset of some facet. Alternatively, any non-empty face of dimension greater than 1 is formed by the convex combination of a set of 1-dimensional faces of the polyhedral cone called \textit{extreme rays} (see \cite{CheYeu12} for more details on the faces of $\Gamma_3$ and their dimension). There are 8 extreme rays of the Shannon region for 3 random variables ($\mathbb{R}_+$ denotes the set of non-negative real numbers):
\begin{align}
R_1 &= \{r_1[1,0,0,1,1,0,1]^\intercal : r_1 \in \mathbb{R}_+\}\\
R_2 &= \{r_2[0,1,0,1,0,1,1]^\intercal : r_2 \in \mathbb{R}_+\}\\
R_3 &= \{r_3[0,0,1,0,1,1,1]^\intercal : r_3\in \mathbb{R}_+\}\\
R_{12} &= \{r_{12}[1,1,0,1,1,1,1]^\intercal : r_{12}\in \mathbb{R}_+\}\\
R_{13} &= \{r_{13}[1,0,1,1,1,1,1]^\intercal : r_{13}\in \mathbb{R}_+\}\\
R_{23} &= \{r_{12}[0,1,1,1,1,1,1]^\intercal : r_{23}\in \mathbb{R}_+\}\\
R_{123} &= \{r_{123}[1,1,1,1,1,1,1]^\intercal : r_{123}\in \mathbb{R}_+\}\\
R_{123'} &= \{r_{123'}[1,1,1,2,2,2,2]^\intercal : r_{123'}\in \mathbb{R}_+\}
\end{align}

The random variables associated with the entropy vectors in the above extreme rays have the following relation:
\begin{itemize}
	\item $R_i= \{\mathbf{h}: Y_j$ and $Y_k$ are constants (degenerates)$\}, i\in \{1,2,3\}$.
	\item $R_{ij}= \{\mathbf{h}: Y_i$ and $Y_j$  are the same and $Y_k$ is a constant$\}, \{i,j\}\subset\{1,2,3\}$.
	\item $R_{123}= \{\mathbf{h}: Y_1,Y_2$ and $Y_3$ are the same$\}$.
	\item $R_{123'}= \{\mathbf{h}: Y_1$, $Y_2$ and $Y_3$ are pairwise independent and each is a function of the remaining two$\}$.\label{0}
\end{itemize}

The extreme rays $R_1$, $R_2$, $R_3$, $R_{12}$, $R_{13}$, $R_{23}$ and $R_{123}$ are plenarily entropic but $R_{123'}$ has non-entropic points. The complete characterization of the entropic points in $R_{123'}$ is as follows.
\begin{theorem}[Zhang and Yeung\cite{ZhaYeu97}]
$\mathbf{h}\in R_{123'}$ is entropic iff \begin{align}
    r_{123'}=\log m, m\in\mathbb{N}.
\end{align}
\end{theorem}

The set of all faces of $\Gamma_3$ containing non-entropic points are formed by convex combinations (i.e., convex hull), denoted $\mathrm{conv}(\cdot)$, of $R_{123'}$ with certain subsets of seven other plenarily entropic extreme rays. Here is a list of such faces (and all the remaining such faces can be obtained by permutation, see \cite{CheYeu12} for details):
\begin{itemize}
	\item  1-dimensional face: $R_{123'}$.
	\item 2-dimensional faces: conv($R_1$, $R_{123'}$), and conv($R_{12}$, $R_{123'}$).
	\item 3-dimensional faces: conv($R_1$, $R_2$, $R_{123'}$), conv($R_{12}$, $R_{13}$, $R_{123'}$), conv($R_{1}$, $R_{12}$, $R_{123'}$) and conv($R_{1}$, $R_{23}$, $R_{123'}$).
	\item 4-dimensional faces: conv($R_1$, $R_2$, $R_3$, $R_{123'}$), conv($R_1$, $R_2$, $R_{12}$, $R_{123'}$), conv($R_1$, $R_2$, $R_{13}$, $R_{123'}$), conv($R_{1}, R_{12}, R_{13}, R_{123'}$), conv($R_1$, $R_{12}$, $R_{23}$, $R_{123'}$) and conv($R_{12}$, $R_{13}$, $R_{23}$, $R_{123}$, $R_{123'}$).
	\item 5-dimensional faces: conv($R_1$, $R_2$, $R_3$, $R_{12}$, $R_{123'}$), conv($R_1$, $R_2$, $R_{12}$, $R_{13}$, $R_{123'}$), conv($R_1$, $R_{2}$, $R_{13}$, $R_{23}$, $R_{123'}$) and conv($R_1$, $R_{12}$, $R_{13}$, $R_{23}$, $R_{123}$, $R_{123'}$).
	\item 6-dimensional faces: conv($R_1$, $R_2$, $R_3$, $R_{12}$, $R_{13}$, $R_{123'}$) and conv($R_1$, $R_2$, $R_{12}$, $R_{13}$, $R_{23}$, $R_{123}$, $R_{123'}$).
\end{itemize}

The complete characterization of the entropic points in 2-dimensional faces is known and is stated in Theorems \ref{thm:2-dim-Matus} and \ref{thm:2-dim-Chen}.
\begin{theorem}[Mat\'{u}\v{s}\cite{Mat06}]\label{thm:2-dim-Matus}
$\mathbf{h}\in$ conv($R_{23}$, $R_{123'}$) is entropic iff
\begin{align}
    r_{123'}+r_{23}\geq \log \lceil k^{r_{123'}}\rceil
\end{align}
where the base of logarithms used in defining the entropy function is $k$.
\end{theorem}
\begin{theorem}[Chen and Yeung \cite{CheYeu12}]\label{thm:2-dim-Chen}
$\mathbf{h}\in$ conv($R_1$, $R_{123'}$) is entropic iff $r_{123'}=\log m, m\in\mathbb{N}$ (in other words, random variables $Y_a, a = 2, 3$ follow the uniform distribution with support size $m$, where $m\in \mathbb{N}$).
\end{theorem}

The following property of the entropy vectors is instrumental to formulate inner bounds in the next section.
\begin{proposition}[Yeung \cite{Yeu08}]\label{prop:1}
		If $\mathbf{h}$ and $\mathbf{h}'$ are in  ${\Gamma}^*_n$, then $\mathbf{h} + \mathbf{h}'$ is also in  ${\Gamma}^*_n$.
\end{proposition}

Only inner and outer bounds for the entropic regions are known for some faces of dimension 3 and 4 containing non-entropic points. But complete characterization is still unknown. 
Inner bounds on all the faces containing non-entropic points can be obtained by Proposition \ref{prop:1} and inner bounds on lower dimensional subfaces whereas, outer bounds on the entropic region for two faces containing non-entropic points have been proved in \cite{HoChaGra12} and are stated here:
\begin{theorem}[Ho, Chan and Grant \cite{HoChaGra12}]\label{prop:3}
If $\mathbf{h}\in$ conv($R_1$, $R_2$, $R_{12}$, $R_{123'}$) is entropic then \begin{align}
    H(Y_a) \geq \log |\mathcal S_3|, a= 1, 2
\end{align}
\end{theorem}
\begin{theorem}[Ho, Chan and Grant \cite{HoChaGra12}]\label{prop:2}
If $\mathbf{h}\in$ conv($R_1$, $R_2$, $R_{123'}$) is entropic then 	\begin{align}
	\max_{y_a\in \mathcal S_a} p(y_a) \leq \min_{y_3\in\mathcal S_3} p(y_3), a\in\{1,2\}.
	\end{align}
\end{theorem}
Note that, Theorem \ref{prop:2} presents a non-entropic inequality which must be followed by the entropic points in the face.
\section{Main Results}\label{sec:Main Results}
\subsection{New Outer Bounds}
An outer bound for the entropy region in the face $\mathbf{h}\in$ conv($R_1$, $R_{23}$, $R_{123'}$) is characterized as follows.
\begin{theorem}\label{th:90th1}
	If $\mathbf{h}\in$ conv($R_1$, $R_{23}$, $R_{123'}$) is entropic then for associated random variables $Y_1, Y_2$ and $Y_3$, the distributions of the  random variables $Y_2, Y_3$ must be the same (and hence, they also have the same support size).
\end{theorem}
\begin{IEEEproof}
	If $\mathbf{h}\in$ conv($R_1$, $R_{23}$, $R_{123'}$) is entropic then for associated random variables $Y_1,Y_2,Y_3$, we have 1) $Y_1$ and $Y_{2}$ are independent, 2) $Y_1$ and $Y_{3}$ are independent and 3) $Y_a, a\in\{2,3\}$ is a function of the remaining two random variables, i.e.,
	\begin{align}
	I(Y_1;Y_3) &= 0,\label{90eq:x1}\\
	I(Y_1;Y_2) &= 0,\label{90eq:x3}\\
	H(Y_2|Y_1, Y_3) &= 0,\label{90eq:x4}\\
	H(Y_3|Y_1, Y_2) &= 0. \label{90eq:x2}
	\end{align}
Now, note that for any $y_1\in \mathcal S_1$ and $y_2\in \mathcal S_2$, by \eqref{90eq:x3} we have, $p(y_1,y_2)>0$. Hence, by \eqref{90eq:x2}, there exists $y_3\in \mathcal S_3$ such that $p(y_1,y_2,y_3)>0.$ Then, by  \eqref{90eq:x1}- \eqref{90eq:x2},
	\begin{align}
	p(y_1, y_2, y_3) &= p(y_1, y_2)\nonumber\\
	&= p(y_1)  p(y_2).\label{90eq:65th1}\\
	p(y_1, y_2, y_3) &= p(y_1, y_3)\nonumber\\
	&=p(y_1)  p(y_3)\label{90eq:65ath1}.
	\end{align}

	By \eqref{90eq:65th1} and \eqref{90eq:65ath1}
	\begin{align}
	\label{90d10th1}p(y_2) &= p(y_3).
	\end{align}
	That is, for any given $y_2\in \mathcal S_2$ there exists  $y_3\in \mathcal S_3$ such that $p(y_2) = p(y_3)$. In other words, $Y_2$ and $Y_3$ must follow the same distribution.
\end{IEEEproof}

In contrast to Theorem \ref{thm:2-dim-Chen}, in which $Y_2$ and $Y_3$ must follow the uniform distribution on supports of the same size, in Theorem \ref{th:90th1} we showed that $Y_2$ and $Y_3$ must follow the same distribution (but not necessarily the uniform distribution). Following is a corollary of Theorem \ref{th:90th1}, describing an entropic equality.
\begin{corollary}
	If a set of random variables $\{Y_1, Y_2, Y_3\}$ satisfies \eqref{90eq:x1}-\eqref{90eq:x2}, then 
	\begin{align}
H(Y_2)=H(Y_3).
	\end{align}
\end{corollary}

A converse-type result (bound) for the entropy region in the face conv($R_1$, $R_{12}$, $R_{123'}$) is characterized in the following theorem. Similar to Theorem \ref{prop:2}, this result too presents a non-entropic inequality which must be followed by the entropic points in the face.
\begin{theorem}\label{th:90}
	If $\mathbf{h}\in$ conv($R_1$, $R_{12}$, $R_{123'}$) is entropic then for associated random variables $Y_1,Y_2$ and $Y_3$,
	\begin{align}
	\label{90re:1}\max_{y_1\in \mathcal S_1} p(y_1) &\leq \min_{y_3\in \mathcal S_3} p(y_3).
	\end{align}
\end{theorem}
\begin{IEEEproof}
	If $\mathbf{h}\in$ conv($R_1$, $R_{12}$, $R_{123'}$) is entropic then for associated random variables $Y_1,Y_2,Y_3$, we have 1) $Y_1$ and $Y_{3}$ are independent, 2) $Y_2$ and $Y_{3}$ are independent and 3) $Y_a, a\in\{2,3\}$ is a function of the remaining two random variables, i.e.,
	\begin{align}
	\label{903'}	I(Y_1;Y_3) &= 0,\\
	\label{904'}	I(Y_2;Y_3) &= 0,\\
	\label{901'}	H(Y_2|Y_1, Y_3) &= 0,\\
	\label{902'}	H(Y_3|Y_1, Y_2) &= 0.
	\end{align}

Let $\mathcal S_a$ denote the support of the random variable $Y_a, a\in\{1, 2, 3\}$. By \eqref{903'} and \eqref{901'}, for any $y_1\in\mathcal S_1$ and $y_3\in \mathcal S_3$, there exists ($y_1, y_2, y_3$) $\in$ $\mathcal{S}_1 \times  \mathcal{S}_2  \times \mathcal{S}_3$ such that $p(y_1, y_2, y_3) > 0$. Then, the probability mass function satisfies
	\begin{align}
	p(y_1, y_2, y_3) &= p(y_1, y_2)\nonumber\\
	&\geq p(y_1)  p(y_2)\label{90eq:65},\\
	p(y_1, y_2, y_3) &\leq p(y_2, y_3)\nonumber\\
	&= p(y_2)  p(y_3). \label{90eq:66}
	\end{align}
Hence,	
	\begin{align}
	\label{90d8}p(y_1) &\leq p(y_3) \Rightarrow \max_{y_1\in \mathcal S_1} p(y_1) \leq \min_{y_3\in\mathcal S_3} p(y_3)
	\end{align}

\end{IEEEproof}

\subsection{Looseness of Known Inner Bounds}

In Theorem \ref{lem: inner1} and \ref{lem: inner2}, we show via explicit construction of distributions that the existing inner bounds for the faces conv($R_{12}$, $R_{23}$, $R_{123'}$) and conv($R_{12}$, $R_{13}$, $R_{23}$, $R_{123}$, $R_{123'}$), respectively, are not tight.

An inner bound for the entropic region in the face formed by conv($R_{12}$, $R_{23}$, $R_{123'}$) can be obtained by Proposition \ref{prop:1} and the characterization of the entropic points in the subfaces conv($R_{12}$, $R_{123'}$) and conv($R_{23}$, $R_{123'}$), see Theorem \ref{thm:2-dim-Matus}. Thus, we obtain the following inner bound on the face conv($R_{12}$, $R_{23}$, $R_{123'}$). 
\begin{align}
\{\mathbf{h}:\mathbf{h}=\mathbf{h}_1+\mathbf{h}_2+\mathbf{h}_3\}\label{eq:3d inner bd 1}
\end{align}
where,
\begin{align}
\mathbf{h}_{1} &=r_{12} [1, 1,0, 1, 1, 1, 1]^\intercal, r_{12} \in \mathbb{R}_+,\\
\mathbf{h}_{2} &=r_{23} [0, 1,1, 1, 1, 1, 1]^\intercal, r_{23} \in \mathbb{R}_+,\\
\mathbf{h}_3 &= r_{123'}[1,1,1,2,2,2,2]^\intercal, r_{123'} \in \mathbb{R}_+,
\end{align}
and at least one of the following two inequalities holds
\begin{align}
r_{12}+r_{123'} &\geq \log \lceil  k^{(r_{123'})}\rceil,\\
r_{23}+r_{123'} &\geq \log \lceil  k^{(r_{123'})}\rceil. \label{eq:3d inner bd 2}
\end{align}

\begin{theorem}\label{lem: inner1}
There exists a distribution such that the corresponding entropy vector 
lies strictly inside the face conv($R_{12}$, $R_{23}$, $R_{123'}$) and strictly outside the inner bound \eqref{eq:3d inner bd 1}-\eqref{eq:3d inner bd 2}.
\end{theorem} 
\begin{IEEEproof}
If $\mathbf{h}\in$ conv($R_{12}$, $R_{23}$, $R_{123'}$) is entropic then for associated random variables $Y_1$, $Y_2$ and $Y_3$, we have 1) each random variable is a function of the other two random variables and 2) $Y_1$ is independent of $Y_3$, i.e., 
	\begin{align}
	\label{90n23}H(Y_3|Y_1, Y_2) &= 0,\\
	\label{90n24}H(Y_2|Y_1, Y_3) &= 0,\\
	\label{90n25}H(Y_1|Y_2, Y_3) &= 0,\\
	\label{90n26}I(Y_1;Y_3)&=0.
	\end{align}

Let $\mathcal S_a=\{0,1\}, a=1,2,3$. The function XNOR satisfies \eqref{90n23}, \eqref{90n24}, \eqref{90n25}. 
	Hence,
	\begin{align}
	\mathrm{Pr}\{Y_2=0|Y_1=0, Y_3=0\}&=0,\\
	\mathrm{Pr}\{Y_2=1|Y_1=0, Y_3=0\}&=1,\\
	\mathrm{Pr}\{Y_2=0|Y_1=0, Y_3=1\}&=1,\\
	\mathrm{Pr}\{Y_2=1|Y_1=0, Y_3=1\}&=0,\\
	\mathrm{Pr}\{Y_2=0|Y_1=1, Y_3=0\}&=1,\\
	\mathrm{Pr}\{Y_2=1|Y_1=1, Y_3=0\}&=0,\\
	\mathrm{Pr}\{Y_2=0|Y_1=1, Y_3=1\}&=0,\\
	\mathrm{Pr}\{Y_2=1|Y_1=1, Y_3=1\}&=1.
	\end{align} 

	Let $p(y_1, y_3)$ be as follows.
	\begin{center}
		\begin{tabular}{ |c|c|c| } 
			\hline
			$Y_1$ & $Y_3$ & $p(y_1, y_3)$  \\ 
			\hline
			0 & 0 & $p$ \\ 
			0 & 1 & $q$  \\
			1 & 0 & $r$	\\
			1 & 1 & $1-(p+q+r)$\\
			\hline
		\end{tabular}
	\end{center}

	From the table of joint distribution of $Y_1, Y_3$ and using the XNOR function, the joint distribution $Y_1, Y_2, Y_3$ is
	\begin{center}
		\begin{tabular}{ |c|c|c|c| } 
			\hline
			$Y_1$ & $Y_2$ &$Y_3$ & $p(y_1, y_2, y_3)$ \\ 
			\hline
			0 & 0 & 0& 0 \\ 
			0 & 0 & 1& $q$ \\  
			0 & 1 & 0& $p$\\
			0 & 1 & 1& 0  \\
			1 & 0 & 0& $r$  \\
			1 & 0 & 1& 0 \\
			1 & 1 & 0& 0  \\
			1 & 1 & 1& $1-(p+q+r)$ \\
			\hline
		\end{tabular}
	\end{center}
	
	Hence, the marginals are as follows. 
\begin{center}
\begin{tabular}{|c|c|c|c|c|c|}
\hline
$Y_1$ & $p(y_1)$  & $Y_2$ & $p(y_2)$  & $Y_3$ & $p(y_3)$  \\ \hline
0     & $p+q$     & 0     & $q+r$     & 0     & $p+r$     \\ 
1     & $1-(p+q)$ & 1     & $1-(q+r)$ & 1     & $1-(p+r)$ \\ \hline
\end{tabular}
\end{center}
	\begin{center}
		\begin{tabular}{ |c|c|c|c| } 
			\hline
			$Y_1$ & $Y_2$ &$p(y_1, y_2)$ & $p(y_1)  p(y_2)$ \\ 
			\hline
			0 & 0& $q$ &$(p+q)  (q+r)$\\ 
			0 & 1& $p$ &$(p+q)  (1-q-r)$\\ 
			1 & 0& $r$ &$(1-p-q)  (q+r)$\\
			1 & 1& $1-(p+q+r)$&$(1-p-q)  (1-q-r)$\\
			\hline
		\end{tabular}
	\end{center}	\begin{center}
		\begin{tabular}{ |c|c|c|c| } 
			\hline
			$Y_1$ & $Y_3$ & $p(y_1, y_3)$ & $p(y_1)  p(y_3)$ \\ 
			\hline
			0 & 0 & $p$ & $(p+q)  (p+r)$ \\ 
			0 & 1 & $q$ & $(p+q)  (1-p-r)$ \\
			1 & 0 & $r$	& $(1-p-q)  (p+r)$\\
			1 & 1 & $1-(p+q+r)$	& $(1-p-q)  (1-p-r)$\\
			\hline
		\end{tabular}
	\end{center}
	\begin{center}
		\begin{tabular}{ |c|c|c|c| } 
			\hline
			$Y_2$ & $Y_3$ &$p(y_2, y_3)$ & $p(y_2)  p(y_3)$ \\ 
			\hline
			0 & 0& $r$ &$(p+r)  (q+r)$\\ 
			0 & 1& $q$ &$(1-p-r)  (q+r)$\\ 
			1 & 0& $p$ &$(p+r)  (1-q-r)$\\
			1 & 1& $1-(p+q+r)$ &$(1-p-r)  (1-q-r)$\\
			\hline
		\end{tabular}
	\end{center}
	
	Using the above mentioned tables of joint distributions, a distribution associated with a point lying strictly inside the face must follow these conditions:
	\begin{itemize}
		\item $(p+q)  (p+r) = p$ (to satisfy \eqref{90n26}).
		\item $(p+q) (q+r) \neq q$ else, the point will lie in the subface formed by conv($R_{23}$, $R_{123'}$).
		\item $(p+r) (q+r)\neq r$ else, the point will lie in the subface formed by conv($R_{12}$, $R_{123'}$).
		\item $(p+q)\neq 0$ else, $Y_1$ will be degenerate and the point will lie in the 1-dimensional subface $R_{23}$ (i.e., corresponding entropy function will be 0 hence, a point will not lie strictly inside the face).
		\item $(p+q)\neq 1$ else, $Y_1$ will be degenerate and the point will lie in the 1-dimensional subface $R_{23}$. 		\item $(q+r)\neq 0$ else, $Y_2$ will be degenerate and the point will be the origin (0-dimensional face).
        \item $(q+r)\neq 1$ else, $Y_2$ will be degenerate and the point will be the origin.
        \item $(p+r)\neq 0$ else, $Y_3$ will be degenerate and the point will be in the 1-dimensional subface $R_{12}$.
        \item $(p+r)\neq 1$ else, $Y_3$ will be degenerate and the point will be in the 1-dimensional subface $R_{12}$.
		\end{itemize} 

	For example, let $q = \frac{1}{8}$ and $r = \frac{1}{24}$. Then to satisfy \eqref{90n26}, let $p$ be as follows.
	\begin{align*}
(p+q)  (p+r) - p &=0\\
\Rightarrow	p &= \frac{10 \pm \sqrt{97}}{24}
	\end{align*}
	 Considering $p = \frac{10 + \sqrt{97}}{24}$, we have

\begin{align*}
		(p+r) (q+r) &= \frac{11 + \sqrt{97}}{144} \neq \frac{1}{24},\\
		(p+q) (q+r) &= \frac{13 + \sqrt{97}}{144} \neq \frac{1}{8},\\
		(p+q) &= \frac{13 + \sqrt{97}}{24} \neq 0,\\
		(p+q) &= \frac{13 + \sqrt{97}}{24} \neq 1,\\
		(q+r) &= \frac{1}{6} \neq 0,\\
		(q+r) &= \frac{1}{6} \neq 1,\\
		(p+r) &= \frac{11 + \sqrt{97}}{24} \neq 0,\\
		(p+r) &= \frac{11 + \sqrt{97}}{24} \neq 1.
\end{align*}

	Thus, for this distribution, it can be verified that the corresponding point lies strictly inside the face.
	The joint probability distribution of the random variables $(Y_1, Y_2)$ and the marginal distributions of $Y_1, Y_2, Y_3$ are
	\begin{center}
		\begin{tabular}{ |c|c|c| } 
			\hline
			$Y_1$ & $Y_3$ & $p(y_1, y_3)$\\ 
			\hline
			0 & 0 & $\frac{10 + \sqrt{97}}{24}$ \\ 
			0 & 1 & $\frac{1}{8}$ \\
			1 & 0 & $\frac{1}{24}$\\
			1 & 1 & $\frac{10 - \sqrt{97}}{24}$ \\
			\hline
		\end{tabular}
	\end{center}
	\begin{center}
\begin{tabular}{|c|c|c|c|c|c|}
\hline
$Y_1$ & $p(y_1)$                    & $Y_2$ & $p(y_2)$      & $Y_3$ & $p(y_3)$                    \\ \hline
0     & $\frac{13 + \sqrt{97}}{24}$ & 0     & $\frac{1}{6}$ & 0     & $\frac{11 + \sqrt{97}}{24}$ \\ 
1     & $\frac{11 - \sqrt{97}}{24}$ & 1     & $\frac{5}{6}$ & 1     & $\frac{13 - \sqrt{97}}{24}$ \\ \hline
\end{tabular}
	\end{center}

Fix $k=2$ (the base of logarithms). Then,
\begin{align*}
H(Y_1) 
 \approx  0.277839,
H(Y_2) 
 \approx  0.649943,\\
{H(Y_3)} 
 \approx  0.561101,
{ H(Y_1, Y_3)}
 \approx 0.838863.
\end{align*}

	 From \eqref{90n23}-\eqref{90n25}, we have 
	 \begin{align*}
	     H(Y_1, Y_2, Y_3)&=H(Y_1, Y_2)= H(Y_1, Y_3)=H(Y_2, Y_3).
	 \end{align*}

	Hence,
	\begin{align}
	\mathbf{h}\approx [0.277839, 0.6499&43, 0.561101, 0.838863,\nonumber\\ &0.838863, 0.838863, 0.838863]^\intercal.
	\end{align}
	From the inner bound expression in \eqref{eq:3d inner bd 1}-\eqref{eq:3d inner bd 2} for the face, $$r_{123'}=H(Y_1, Y_2)- H(Y_2)\approx 0.188920.$$ 

Moreover,
		\begin{align}
	\label{90amch:1'}	r_{12}+r_{123'}=H(Y_1) \approx 0.277839<0.3,\\
	\label{90amch:1}	r_{23}+r_{123'}=H(Y_3) \approx 0.561101<0.6,\\
	\label{90amch:2}	1=\lceil \log_2 2^{0.1}\rceil\leq\lceil \log_2 2^{r_{123'}}\rceil\leq \lceil \log_2 2^{0.2}\rceil= 1.
	\end{align}
	From \eqref{90amch:1'}-\eqref{90amch:2}, 
	\begin{align}
	    r_{12}+r_{123'} \ngeqslant \lceil \log_2 2^{r_{123'}} \rceil,\\
	    r_{23}+r_{123'} \ngeqslant \lceil \log_2 2^{r_{123'}} \rceil.
	\end{align}
	Therefore, this point does not lie in the inner bound. 
\end{IEEEproof}

An inner bound for the entropic region in the face formed by conv($R_{12}$, $R_{13}$, $R_{23}$, $R_{123}$, $R_{123'}$) can be obtained by Proposition \ref{prop:1} and the characterization of the entropic points in the subfaces conv($R_{12}$, $R_{123'}$), conv($R_{13}$, $R_{123'}$) and conv($R_{23}$, $R_{123'}$), see Theorem \ref{thm:2-dim-Matus}. Thus we obtain the following inner bound for the face conv($R_{12}$, $R_{13}$, $R_{23}$, $R_{123}$, $R_{123'}$).  
\begin{align}
\{\mathbf{h}:\mathbf{h}=\mathbf{h}_1+\mathbf{h}_2+\mathbf{h}_3+\mathbf{h}_4+\mathbf{h}_5\}\label{eq:3d inner bd 1'}
\end{align}
where,
\begin{align}
\mathbf{h}_{1} &=r_{12} [1, 1,0, 1, 1, 1, 1]^\intercal, r_{12} \in \mathbb{R}_+,\\
\mathbf{h}_{2} &=r_{13} [1, 0,1, 1, 1, 1, 1]^\intercal, r_{13} \in \mathbb{R}_+,\\
\mathbf{h}_{3} &=r_{23} [0, 1,1, 1, 1, 1, 1]^\intercal, r_{23} \in \mathbb{R}_+,\\
\mathbf{h}_{4} &=r_{123} [1, 1,1, 1, 1, 1, 1]^\intercal, r_{123} \in \mathbb{R}_+,\\
\mathbf{h}_5 &= r_{123'}[1,1,1,2,2,2,2]^\intercal, r_{123'} \in \mathbb{R}_+,
\end{align}
and at least one of the following three inequalities holds
\begin{align}
r_{12}+r_{123'} &\geq \log \lceil  k^{(r_{123'})}\rceil,\\
r_{13}+r_{123'} &\geq \log \lceil  k^{(r_{123'})}\rceil,\\
r_{23}+r_{123'} &\geq \log \lceil  k^{(r_{123'})}\rceil. \label{eq:3d inner bd 2'}
\end{align}
\begin{theorem}\label{lem: inner2}
There exists a distribution such that the corresponding entropy vector lies strictly inside the face conv($R_{12}$, $R_{13}$, $R_{23}$, $R_{123}$, $R_{123'}$) and strictly outside the inner bound \eqref{eq:3d inner bd 1'}-\eqref{eq:3d inner bd 2'}. 
\end{theorem}

\begin{IEEEproof}
If $\mathbf{h}\in$ conv($R_{12}$, $R_{13}$, $R_{23}$, $R_{123}$, $R_{123'}$) is entropic then for associated random variables $Y_1$, $Y_2$ and $Y_3$, we have (1) each random variable is a function of the remaining two random variables, i.e., 
	\begin{align}
	\label{90n13}H(Y_3|Y_1, Y_2) &= 0,\\
	\label{90n14}H(Y_2|Y_1, Y_3) &= 0,\\
	\label{90n16}H(Y_1|Y_2, Y_3) &= 0.
	\end{align}

Let $\mathcal S_{a}=\{0,1\}, a=1,2,3$. Using the approach similar to that in the proof of Theorem \ref{lem: inner1}, we let $Y_3$ be the XNOR function of $Y_1, Y_2$ and find the following joint distribution and marginals such that the corresponding entropic vector is strictly inside the face conv($R_{12}$, $R_{13}$, $R_{23}$, $R_{123}$, $R_{123'}$). 
	\begin{center}
		\begin{tabular}{ |c|c|c| } 
			\hline
			$Y_1$ & $Y_2$ & $p(y_1, y_2)$\\ 
			\hline
			0 & 0 & $\frac{1}{3}$ \\ 
			0 & 1 & $\frac{1}{8}$ \\
			1 & 0 & $\frac{1}{24}$\\
			1 & 1 & $\frac{1}{2}$\\
			\hline
		\end{tabular}
	\end{center}
	\begin{center}
\begin{tabular}{|c|c|c|c|c|c|}
\hline
$Y_1$ & $p(y_1)$        & $Y_2$ & $p(y_2)$      & $Y_3$ & $p(y_3)$      \\ \hline
0     & $\frac{11}{24}$ & 0     & $\frac{3}{8}$ & 0     & $\frac{1}{6}$ \\ 
1     & $\frac{13}{24}$ & 1     & $\frac{5}{8}$ & 1     & $\frac{5}{6}$ \\ \hline
\end{tabular}
	\end{center}
	
Fix $k=2$. 
Then, 
$H(Y_1) \approx 0.994984$, $ H(Y_2) \approx 0.954434$, $H(Y_3) \approx 0.650022$, $H(Y_1, Y_2)\approx1.594360$
and
	\begin{align*}
	    H(Y_1, Y_2)
	    = H(Y_1, Y_3)= H(Y_2, Y_3)&=H(Y_1, Y_2, Y_3)
	\end{align*}	
implies
	\begin{align}
	\mathbf{h}\approx [0.994984, 0.9544&34, 0.650022, 1.594360,\nonumber\\ &1.594360, 1.594360, 1.594360]^\intercal.
	\end{align}

	 From the inner bound for the given face in \eqref{eq:3d inner bd 1'}-\eqref{eq:3d inner bd 2'}, 
	 \begin{align}
	 r_{23}+r_{123'}=H(Y_1, Y_2)-H(Y_1)&\approx 0.599376,\label{eq:1}\\
	 r_{13}+r_{123'}=H(Y_1, Y_2)-H(Y_2)&\approx 0.639926,\label{eq:2}\\
	 r_{12}+r_{123'}= H(Y_1, Y_2)-H(Y_3)&\approx 0.944338,\label{eq:3}
	 \end{align}
	 and
	 \begin{align}
	 r_{123'}-r_{123}&=2H(Y_1, Y_2)-H(Y_1)-H(Y_2)-H(Y_3)\nonumber\\
	 &\approx 0.589280. \label{eq:4}
	 \end{align}
	 Thus, from \eqref{eq:1} and \eqref{eq:4}, $0.58 <r_{123'}<0.60$. 
	 Moreover,
	\begin{align}
	\label{90amch:2new}	\lceil \log_2 2^{r_{123'}}\rceil\geq \lceil \log_2 2^{0.58}\rceil&= 1,\\
	\label{90amch:2new2}\lceil \log_2 2^{r_{123'}}\rceil\leq \lceil \log_2 2^{0.60}\rceil &= 1,
	\end{align}
and hence,  
	\begin{align}
	    r_{12}+r_{123'} \ngeqslant \lceil \log_2 2^{r_{123'}}\rceil,\\ r_{13}+r_{123'} \ngeqslant \lceil \log_2 2^{r_{123'}}\rceil,\\
	    r_{23}+r_{123'} \ngeqslant \lceil \log_2 2^{r_{123'}}\rceil. 
	\end{align}

	Therefore, this point does not lie in the inner bound for the given range of values of $r_{123'}$. Thus we have shown existence of a distribution such that corresponding entropy vector lies strictly inside the face conv($R_{12}$, $R_{13}$, $R_{23}$, $R_{123}$, $R_{123'}$) and strictly outside the inner bound \eqref{eq:3d inner bd 1'}-\eqref{eq:3d inner bd 2'}.
\end{IEEEproof}

\section{Conclusion}\label{sec:Conclusion}
Outer bounds (converse-type results) for the entropy region in two faces are presented. We also showed that known inner bounds for two faces are loose. The approach used can be applied to study tightness of known inner bounds or to obtain distributions corresponding to entropic points in a constrained entropy region. Since the known inner bounds for two faces are not tight (Theorem \ref{lem: inner1} and \ref{lem: inner2}), a natural future direction is to characterize better inner bounds for the entropy regions in these faces by utilizing the XNOR (or XOR) function.

\section*{Acknowledgment}
This work is supported by SERB, DST, Government of India, under Extra Mural Scheme SB/S3/EECE/265/2016. We thank the reviewers for the comments and suggestions.
\bibliographystyle{ieeetr}
\bibliography{network}

\end{document}